\documentclass[sigconf]{acmart}

\AtBeginDocument{%
  }

\setcopyright{acmcopyright}
\copyrightyear{2023}
\acmYear{2023}
\acmDOI{XXXXXXX.XXXXXXX}

\acmConference[XX]{XX}{XX}{XX}
\acmBooktitle{XX,X,X}
\acmPrice{15.00}
\acmISBN{XXX-X-XXXX-XXXX-X/23/08}

\usepackage{amsmath}
\usepackage{multirow}
\usepackage{booktabs}
\usepackage{array}
\usepackage{subfigure}
\usepackage{amssymb}
\usepackage{algorithmic}
\usepackage{algorithm}
\usepackage{verbatim}
\usepackage{bm}     
\usepackage{enumitem}       
\begin{document}

\title{Learning from All Sides: Diversified Positive Augmentation via Self-distillation in Recommendation}


\author{Chong Liu, Xiaoyang Liu, Ruobing Xie, Lixin Zhang, Feng Xia, Leyu Lin}
\email{nickcliu@tencent.com}
\affiliation{%
  \institution{Tencent}
  \country{China}
  }

\renewcommand{\shortauthors}{Chong Liu et al.}

\begin{abstract}
Personalized recommendation relies on user historical behaviors to provide user-interested items, and thus seriously struggles with the data sparsity issue. A powerful positive item augmentation is beneficial to address the sparsity issue, while few works could jointly consider both the accuracy and diversity of these augmented training labels. In this work, we propose a novel model-agnostic Diversified self-distillation guided positive augmentation (DivSPA) for accurate and diverse positive item augmentations. Specifically, DivSPA first conducts three types of retrieval strategies to collect high-quality and diverse positive item candidates according to users' overall interests, short-term intentions, and similar users. Next, a self-distillation module is conducted to double-check and rerank these candidates as the final positive augmentations. Extensive offline and online evaluations verify the effectiveness of our proposed DivSPA on both accuracy and diversity. DivSPA is simple and effective, which could be conveniently adapted to other base models and systems. Currently, DivSPA has been deployed on multiple widely-used real-world recommender systems.
\end{abstract}




\maketitle

\section{Introduction}

Personalized recommendation aims to recommend customized items that users are interested in to satisfy users' diverse preferences. In practice, users' positive historical behaviors (e.g., click, dwell time, purchase) are essential sources to capture user preferences, while these user positive feedbacks are relatively sparse considering the million-level item candidates in the overall item corpora. The intrinsic data sparsity issue of recommendation seriously limits the optimization of models. 

Nowadays, lots of efforts concentrate on alleviating the positive data sparsity issue. Some works introduce external information from other domains (cross-domain recommendation \cite{li2020ddtcdr,zhu2022personalized}), behaviors (multi-behavior recommendation \cite{jin2020multi,wu2022multi}), and models (recommendation pre-training \cite{zeng2021knowledge,liu2023graph}) for more sufficient training. With the thriving of self-supervised learning (SSL) widely verified in various fields, some works also attempt to build certain augmentations to bring in additional training signals without external information. The majority of these works focus on certain user augmentations based on their behaviors and attributes \cite{zhou2020s3,wu2022disentangled}. In contrast, other works build item augmentations from another aspect \cite{yao2021self,xie2023multi}. \cite{liu2023ufnrec} focuses on filtering false negative samples and directly converts high-quality items from randomly selected negative samples into positive samples supervised by self-distillation.
There are two challenges in positive item augmentations of recommendation: (1) the positive item augmentation should be accurate to enable better training. (2) The sources of augmented items should be diverse to avoid possible filter bubbles and homogenization issues.

In this work, we attempt to design an accurate, diversified, and model-agnostic framework for positive item augmentation in recommendation. To address the above challenges, we propose a novel Diversified self-distillation guided positive augmentation (DivSPA). First, for each user-item interaction, we conduct three types of retrieval strategies to fast select high-quality items as possible candidates for our positive augmentation. Specifically, we have (a) \emph{u2i retrieval}, which aims to extract users' cumulative interests in both short-term and long-term behaviors, (b) \emph{i2i retrieval}, which concentrates more on users' current intentions or occasional interests of the interacted items, and (c) \emph{u2u2i retrieval}, which focuses on more generalized interests voted by similar users. Next, we adopt a self-distillation mechanism to double-check the qualities of all retrieved item candidates from different types of retrieval strategies and generate the final high-quality and diversified positive item augmentations. These positive augmentations are then added to the train set with weights for more comprehensive training.

In experiments, we conduct extensive evaluations to verify the effectiveness of our DivSPA on both accuracy and diversity. We also conduct an online A/B test to confirm the power of DivSPA in real-world scenarios. Currently, DivSPA has been deployed online in a widely-used recommender system, affecting millions of users. The contributions are concluded as follows:
\begin{itemize}
    \item We propose an effective and universal DivSPA framework for diversified positive item augmentation in recommendation. To the best of our knowledge, this is the first attempt to combine multiple high-quality sources for more accurate and diversified positive augmentation via self-distillation.
    \item We have verified our model in both offline datasets and online evaluations. Currently, DivSPA has been deployed in a widely-used DivSPA has been deployed online in a widely-used recommender system.
\end{itemize}

\section{Related Works}
\noindent
\textbf{Two-tower Recommendation.}
The two-tower architecture is widely used in the recommendation system, which typically trains separate representations for users and items using two towers \cite{covington2016deep}. Based on the two-tower architecture, some sequence models \cite{sasrec, gru4rec, causerec, bert4rec, CT4Rec} have been proposed that focus on modeling the user's historical behavior sequence in the user tower. Recently, contrastive learning \cite{yao2021self, CLRec, CL4SRec, chen2022intent, nie2022mic} has achieved significant performance improvements in the recommendation system's matching tasks. In addition, multi-interest user modeling \cite{cen2020controllable, li2019multi} has also achieved promising results, attracting the attention of many researchers.

\noindent
\textbf{Data augmentation.}
Many researchers have applied data augmentation methods \cite{wei2019eda, he2020momentum, oord2018representation} to improve model performance. For CV tasks, SimCLR \cite{chen2020simple} conducts contrastive learning based on augmentations of the same instances. Integrating data augmentation and contrastive learning \cite{yao2021self, CL4SRec, nie2022mic, wu2022disentangled, xie2023multi} is also widely used for recommendation tasks. Besides, UFNRec \cite{liu2023ufnrec} generates augmented positive samples upon false negatives. In this paper, we achieve diversified positive augmentation via self-distillation.

\section{Methodology}

\subsection{Preliminary}
\noindent
\textbf{Notions.}
We first briefly introduce some necessary notations in our method. In the retrieval task, user features $x_u$ (i.e., user profiles and user behaviors) and item features $x_v$ (i.e., item profiles and item interactions) are sparse inputs, and the label $y$ represents whether the user clicks the item. Then, a base model outputs the user representation $\bm{u}=\mathrm{Enc}(\bm{x}_u)$ and the item representation $\bm{v}=\mathrm{Enc}(\bm{x}_v)$. Since our method has no restriction on the model structure, the base model can be a simple deep neural network, a transformer structure, etc. For each positive pair $(\bm{u},\bm{v}^+)$, we randomly sample $n$ negative items $(\bm{v_1}^-,\dots,\bm{v_n}^-)$ and compute the semantic relevance scores via a softmax function as the training loss:
\begin{equation}
    \begin{split}
    \mathcal{L}_{ori}(\bm{u,v};\omega) = -log(
    \frac{exp(\bm{u v^+})}{exp(\bm{u v^+}) +  \sum\limits_{1\le j \le n}exp(\bm{uv_j^-})})
    \label{func:basic_loss}
    \end{split}
\end{equation}

\begin{figure}[t]
    \centering
    \includegraphics[width=0.5\textwidth]{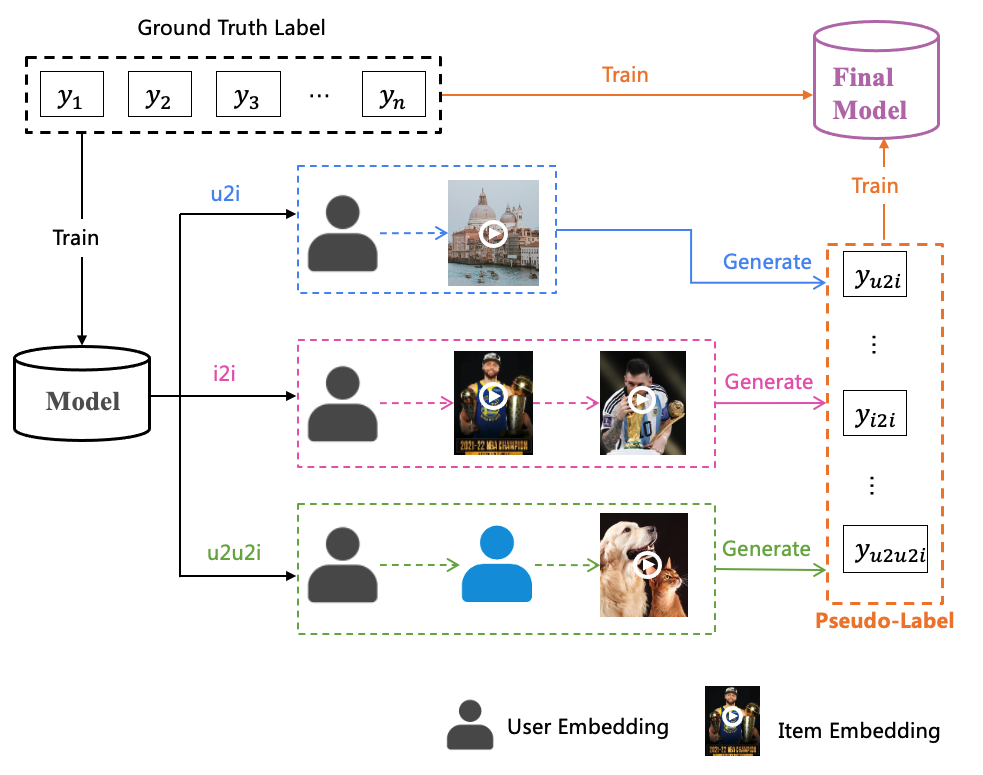}
    \caption{The model structure of DivSPA. Besides the original samples, DivSPA conducts three retrieval strategies to extend diverse positive candidates. The self-distillation module is conducted to double-check the positive augmentations. }  \label{img:main}
    \end{figure}

\noindent
\textbf{Overall Framework.}
Fig. \ref{img:main} shows the overall framework of DivSPA. 
Firstly, we train a base model, i.e., YoutubeDNN, with original instances to obtain primary representations $\bm{u}$ and $\bm{v}^+$. Then, we extend positive samples via three augmentation approaches, i.e., \emph{u2i}, \emph{i2i}, and \emph{u2u2i}. Finally, to alleviate extra noise and guarantee accuracy, we train DivSPA with a mix-up loss to integrate augmented positive samples and original instances. 

\begin{table*}[t]
\caption{Results on two datasets with five baselines. The improvements of DivSPA are significant over all baselines ( p<0.05).}
\center
\begin{tabular}{l|l|p{1.4cm}<{\centering}p{1.4cm}<{\centering}p{1.4cm}<{\centering}
p{1.45cm}<{\centering}p{1.45cm}<{\centering}p{1.6cm}<{\centering}p{1.6cm}<{\centering}}
\toprule
Dataset                 & Model                     & HR@50   & HR@100  & HR@200  & NDCG@50 & NDCG@100 & NDCG@200 \\
\midrule
\multirow{6}{*}{Video}  & FM                        & 0.2059 & 0.2890 & 0.3778 & 0.1050 & 0.1207  & 0.1617  \\
                        & Youtube                  & 0.2204 & 0.3018 & 0.3926 & 0.1109 & 0.1236  & 0.1671  \\
                        & AutoInt                   & 0.2168 & 0.3037 & 0.3920 & 0.1143 & 0.1251  & 0.1671  \\
                        & DeepFM                    & 0.2238 & 0.3041 & 0.3951 & 0.1122 & 0.1243  & 0.1655  \\
                        & RFM                       & 0.2254 & 0.3063 & 0.3979 & 0.1150 & 0.1285  & 0.1707  \\
                        & DivSPA & \textbf{0.2304} & \textbf{0.3114} & \textbf{0.4035} & \textbf{0.1181} & \textbf{0.1310}  & \textbf{0.1735}  \\
                        \midrule
\multirow{6}{*}{Vfeeds} & FM                        & 0.1724 & 0.2915 & 0.4215 & 0.1014 & 0.1290  & 0.1696  \\
                        & Youtube                  & 0.1873 & 0.3064 & 0.4509 & 0.1067 & 0.1379  & 0.1800  \\
                        & AutoInt                   & 0.1900 & 0.3081 & 0.4517 & 0.1090 & 0.1373  & 0.1815  \\
                        & DeepFM                    & 0.1850 & 0.3104 & 0.4551 & 0.1087 & 0.1399  & 0.1810  \\
                        & RFM                       & 0.1932 & 0.3119 & 0.4549 & 0.1109 & 0.1415  & 0.1827  \\
                        & DivSPA & \textbf{0.1961} & \textbf{0.3180} & \textbf{0.4594} & \textbf{0.1145} & \textbf{0.1451}  & \textbf{0.1869}   \\
                       \bottomrule
\end{tabular}
\label{tab:main}
\end{table*}
\subsection{Positive Samples Self-Distillation}
\label{sec.positive_augmentation}
We conduct three strategies to collect high-quality and diverse positive items. We first train the base model with $\mathcal{L}_{ori}$ as Eq.(\ref{func:basic_loss}) to obtain user and item representations. Then, we apply semantic relevance scores, i.e., $s(u,v)=cosine(\bm{uv})$, to extend positive samples.

\noindent
\textbf{Extension via Self-Distillation.}
We explore potential positive items via three self-distillation methods, namely \emph{u2i, i2i} and \emph{u2u2i} for short. Concretely, for each positive sample pair $(\bm{u},\bm{v}^+)$:
\begin{itemize}
\item \emph{u2i.}
We compute the top $k$ item representations $\bm{v}_{u2i}^+$ via their semantic relevance scores $s(u,v_{u2i})$ with the user representation $\bm{u}$, and treat $\bm{v}_{u2i}^+$ as the newly explored positive items for the user u.
\item \emph{i2i.}
We regard the current positive item $\bm{v}^+$ as the user's short-term intention, and then extend top $k$ potential positive items $\bm{v}_{i2i}^+$ based on semantic relevance scores $s(v^+,v_{i2i})$ with $\bm{v}^+$.
\item \emph{u2u2i.}
We retrieve the top $k_u$ user representations $\bm{u}_{u2u}^+$ based on $s(u,u_{u2u})$ with the user representation $\bm{u}$, and then regard the real positive items of users $\bm{u}_{u2u}^+$ as $\bm{v}_{u2u2i}^+$ for $\bm{u}$. 
\end{itemize}
Different from current collaborative filtering methods that utilize the relations of the item and user representations to achieve additional retrieval results, we apply the representations to extend potential positive samples for training.
The above three methods combine the real clicks with the item and user representations from different perspectives. \emph{u2i} directly relies on the item and user representations to retrieve potential positive items, considering both short-term and long-term user interests. \emph{i2i} only depends on current item representations to extend samples, which concentrates more on the user's current or occasional interests. As for \emph{u2u2i}, it contains the relations between user representations \emph{u2u} and real clicks \emph{u2i}, which brings more undiscovered information and relations than other methods.

\noindent
\textbf{Positive Sampling.} After exploring the top $k$ potential positive items from each self-distillation method, we separately sample $m$ items from the top $k$ to reduce the dependence upon semantic relevance scores and improve model robustness. We introduce three sampling methods:
(a) \emph{uniform sampling}, which samples each item with the probability $1/k$.
(b) \emph{importance sampling}, which applies the corresponding normalized relevance scores as the importance to sample items.
(c) \emph{beta sampling}, with the distribution $beta(\alpha,\alpha)$ to randomly sample items.
Thus, we extend the training distribution via self-distillation and bring model-agnostic information. Also, the augmented positive items can be diverse to avoid possible filter bubbles and homogenization issues.
In this paper, we set $m=3$ for each self-distillation approach.

\subsection{Mix-Up Loss}
After obtaining virtual pairs from the vicinity distribution of training representations, we apply convex combinations of these samples instead of directly training them to regularize the model.
A direct way to mix up these virtual samples is to combine their representations as $\Tilde{\bm{v}}=\bm{v}+\beta\sum_\mathcal{SD} \frac{s(u,v_i)}{\sum_{j=1}^{m} s(u,v_j)}\bm{v_i}$,
with $\mathcal{SD}=\left\{\emph{u2i},\emph{i2i},\emph{u2u2i}
\right\}$. Here, $(\bm{u},\bm{v})$ is the original positive pair, and $\bm{v_i}$ is the $i-th$ extended positive item. $\beta\ge0$ is a hyper-parameter to control the influence of mix-up.
Then, the linear combination $(\bm{u},\Tilde{\bm{v}})$ can be trained as :
\begin{equation}
    \begin{split}
    \mathcal{L}(\bm{u,\Tilde{v}};\omega) = -log(
    \frac{exp(\bm{u\Tilde{v}^+})}{exp(\bm{u \Tilde{v}^+}) +  \sum\limits_{1\le j \le n}exp(\bm{\bm{u}v_j^-})})
    \label{func:mixup_pre}
    \end{split}
\end{equation}
However, \cite{zhang2017mixup} finds that the convex combinations of more than three samples can dramatically increase the computation costs while providing limited performance gain.
Besides, such linear combined virtual representations can sometimes be affected by certain features and provide misleading information to the model training.
Therefore, we consider mixing up samples in the aspect of the output space instead of the representation space. Concretely, we finally design the mix-up loss as:
\begin{equation}
    \begin{split}
    \mathcal{L}_{mixup} = \mathcal{L}_{ori}(\bm{u,v};\omega) + \beta\sum_\mathcal{SD} \frac{s(u,v_i)}{\sum_{i=1}^{m} s(u,v_i)} \mathcal{L}_{ori}(\bm{u,v_i};\omega)
    \label{func:mixup_loss}
    \end{split}
\end{equation}
We directly train DivSPA with the mix-up loss, which can more comprehensively train high-quality and diversified positive item augmentation with weights. Generally, DivSPA considers both accuracy and diversity for positive augmentation via self-distillation.

\begin{table}[]
\caption{Detailed statistics of large-scale industrial datasets.}
\label{tab:dataset}
\center
\small
\begin{tabular}{l|ccccc}
\toprule
Dataset &\# user&\# item&\# click&\# fields&\# features \\
\midrule
Video &11.62M&0.42M&25.34M&57&41.2M\\
Vfeeds &1.34M&0.27M&13.58M&50&20.6M\\
\bottomrule
\end{tabular}
\vspace{-0.45cm}
\end{table}

\section{Experiments}
\subsection{Datasets and Settings}
\label{sec.datasets}
\noindent \textbf{Datasets.}
Since few public datasets contain industrial-scale items with detailed features, we build two large-scale datasets from a widely-used recommender system \textit{Topstory}.
The \textit{Video} dataset contains 11.62 million users, 25.34 million click instances, and nearly 41.2 million features. The \textit{Vfeeds} dataset is collected from an immersive video-streaming scene and includes almost 1.34 million users, 13.58 million click instances, and 20.6 million features.
All instances are split into train and test sets in chronological order. The detailed statistics are presented in Table~\ref{tab:dataset}.

\noindent \textbf{Experimental Settings.}
We optimize all models using Adam with a learning rate of $0.001$ and a batch size of $256$. The embedding size of all features is set to $32$. Our method applies a two-tower architecture and we conduct a grid search to select hyper-parameters.

\subsection{Offline Evaluation} 
\label{sec.competitors}
\noindent \textbf{Competitors.}
We choose several high-performing models widely used in the real-world matching stage as baselines for comparison:
(a) \textbf{FM \cite{rendle2010factorization}}, which utilizes latent vectors to model second-order feature interactions.
(b) \textbf{YoutubeDNN \cite{covington2016deep}}, which learns high-order feature interactions by adding DNN structures into a two-tower architecture.
(c) \textbf{DeepFM \cite{guo2017deepfm}}, which integrates DNN and FM models within the two-tower architecture to construct user and item representations separately.
(d) \textbf{AutoInt \cite{song2019autoint}}, which employs a self-attention mechanism to automatically model high-order feature interactions with good model explainability.
(e) \textbf{RFM \cite{yao2021self}}, an item-based collaborative learning model that effectively addresses the challenge of inadequate training for long-tail items to enhance model performance.
We evaluate these methods by HR\({@k}\) and NDCG\({@k}\) with $k=\left\{50,100,200\right\}$ due to the large scale of datasets.

\begin{table*}[t]
\caption{Comparison of the performance between DivSPA and UFNRec.}
\label{tab:ufnrec}
\center
\small
\begin{tabular}{l|p{1.45cm}<{\centering}p{1.45cm}<{\centering}p{1.45cm}<{\centering}
p{1.45cm}<{\centering}p{1.45cm}<{\centering}p{1.5cm}<{\centering}p{1.5cm}<{\centering}}
\toprule
Video & HR@50 & HR@100 & HR@200 & NDCG@50 &  NDCG@100 & NDCG@200 \\
\midrule
\ \ DivSPA(Youtube) & 0.2304 & 0.3114 & 0.4035 & 0.1181 & 0.1310 & 0.1735 \\
\ \ UFNRec & 0.2275  & 0.3078  & 0.3968  & 0.1163  & 0.1276  & 0.1691  \\
\ \ DivSPA(UFNRec) &  \textbf{0.2358}  & \textbf{0.3143}  & \textbf{0.4069}  & \textbf{0.1211}  & \textbf{0.1344}  & \textbf{0.1751}  \\

\bottomrule
\end{tabular}
\end{table*}

\begin{table*}[t]
\caption{Results of ablation study. All improvements of different components are significant (t-test with p$<$0.05).}
\label{tab:ablation}
\center
\small
\begin{tabular}{l|p{1.45cm}<{\centering}p{1.45cm}<{\centering}p{1.45cm}<{\centering}
p{1.45cm}<{\centering}p{1.45cm}<{\centering}p{1.5cm}<{\centering}p{1.5cm}<{\centering}}
\toprule
Video & HR@50 & HR@100 & HR@200 & NDCG@50 &  NDCG@100 & NDCG@200 \\
\midrule
DivSPA(Youtube) & \textbf{0.2304} & \textbf{0.3114} & \textbf{0.4035} & \textbf{0.1181} & \textbf{0.1310} & \textbf{0.1735} \\
\midrule
\ \ w/o u2i & 0.2292 & 0.3090 & 0.4009 & 0.1175 & 0.1293 & 0.1714 \\
\ \ w/o i2i & 0.2259 & 0.3064 & 0.3978 & 0.1150 & 0.1264 & 0.1691 \\
\ \ w/o u2u2i & 0.2262 & 0.3071 & 0.3985 & 0.1159 & 0.1271 & 0.1700 \\
\ \ w/o all & 0.2204 & 0.3018 & 0.3926 & 0.1109 & 0.1236 & 0.1671 \\
\bottomrule
\end{tabular}
\end{table*}


\noindent \textbf{Main Results.}
As shown in Table \ref{tab:main}, we compare DivSPA with other matching models to evaluate the effectiveness of DivSPA. The results show that RFM achieves the best performance among all baseline methods, by applying contrastive learning techniques to obtain better long-tail item embeddings. Despite our backbone being only YoutubeDNN,  it is obvious that DivSPA outperforms all baselines on all metrics, indicating that our diversified positive augmentation method can improve the model performance to surpass that of more complex models.

\noindent \textbf{More Data Augmentation.}
UFNRec \cite{liu2023ufnrec} is a SOTA data augmentation model that explores false negative samples and utilizes self-distillation to eliminate noisy samples and improve recommendation quality. Thus, we compare DivSPA with UFNRec to evaluate the university and effectiveness of DivSPA. As shown in Table \ref{tab:ufnrec}, DivSPA with YoutubeDNN outperforms UFNRec in all metrics, which proves the efficiency of DivSPA. Since UFNRec achieves data augmentation in the aspect of utilizing false negatives, combining DivSPA and UFNRec can further improve the performance of DivSPA with YoutubeDNN.
\subsection{Online A/B Test}
\label{sec.online_evaluation}
To validate the effectiveness of DivSPA in real-world recommendation scenarios, we conduct an online test on a video stream scenario in the recommendation system \textit{Topstory}. The video scenario has already deployed multiple recall methods, including the YoutubeDNN model recall, user geographic recall, Graph model recall, attribute recall, cross-domain recall, i2i recall, and collaborative filtering recall. We conduct two experiments, both lasting for four days on the online platform and affecting over one million users. The two main metrics we focus on are: (a) average play numbers per capita (APN), and (b) video completion rate (VCR).
In the first experiment, we replace the YoutubeDNN recall with DivSPA, resulting in an improvement of +1.604\% in APN and +3.043\% in VCR. In the second experiment, we replace the item embeddings from YoutubeDNN with those generated by DivSPA for i2i recall,  which brings a remarkable increase of +2.316\% in APN and +2.416\% in VCR. All the online improvements are significant with p$<$0.01. These results demonstrate the power of our method to provide users with more satisfying items and increase their interest in our video stream.
\subsection{Ablation Study}
\label{sec.ablation_study}
As shown in Table \ref{tab:ablation}, we conduct four versions of our method on the \textit{Video} dataset to evaluate the effectiveness of different components in DivSPA. The experiment results indicate that:
(1) While the different versions of our method still outperform the baseline YoutubeDNN, our final model DivSPA achieves significantly better performance than all of the other versions.
(2) Since the \textit{i2i} and  \textit{u2u2i} modules introduce more unexpected information, removing these two modules results in a greater decrease in performance compared to the experiment of removing \textit{u2i}.




\subsection{Explorations on Diversity}
\label{sec.Diversity}
Diversity is a significant challenge in real-world recommendation systems and essential to overcome filter bubbles. To evaluate the diversity of recommendation results, we compare the number of distinct items retrieved by DivSPA and YoutubeDNN in the top 100, top 500, and top 1000 results for 112,000 users. As shown in Table \ref{tab:ufnrec}, DivSPA retrieves over three times more items than YoutubeDNN on all metrics, which indicates that DivSPA can recall more long-tail items and brings better diversity. 
Thus, DivSPA can simultaneously improve accuracy and diversity by comprehensively involving positive item augmentation in model training.

\begin{table}[]
\caption{Comparison of the diversity between DivSPA and YoutubeDNN.}
\label{tab:diversity}
\center
\small
\begin{tabular}{l|p{1.45cm}<{\centering}p{1.45cm}<{\centering}p{1.45cm}<{\centering}
p{1.45cm}<{\centering}p{1.45cm}<{\centering}p{1.5cm}<{\centering}p{1.5cm}<{\centering}}
\toprule
 recalled items  & top-100 & top-500 & top-1000  \\
\midrule
\ \ Youtube   & 9134 & 18910 & 43566   \\
\ \ DivSPA(Youtube) & 51702  & 93675  & 167481    \\
\ \ improvement &  566\%  & 495\%  & 384\%    \\
\bottomrule
\end{tabular}
\end{table}
\section{Conclusion and Future Work}
In this work, we propose a novel self-distillation guided positive augmentation for accurate and diverse positive item augmentations.
We have deployed DivSPA on multiple widely-used real-world recommender systems. In the future, we will explore more positive augmentation methods and theoretically analyze the pros and cons of our method.

\bibliographystyle{ACM-Reference-Format}
\bibliography{references}

\begin{thebibliography}{}

\bibitem[Cen et~al., 2020]{cen2020controllable}
Cen, Y., Zhang, J., Zou, X., Zhou, C., Yang, H., and Tang, J. (2020).
\newblock Controllable multi-interest framework for recommendation.
\newblock In {\em Proceedings of the 26th ACM SIGKDD International Conference
  on Knowledge Discovery \& Data Mining}, pages 2942--2951.

\bibitem[Chen et~al., 2020]{chen2020simple}
Chen, T., Kornblith, S., Norouzi, M., and Hinton, G. (2020).
\newblock A simple framework for contrastive learning of visual
  representations.
\newblock In {\em Proceedings of ICML}.

\bibitem[Chen et~al., 2022]{chen2022intent}
Chen, Y., Liu, Z., Li, J., McAuley, J., and Xiong, C. (2022).
\newblock Intent contrastive learning for sequential recommendation.
\newblock In {\em Proceedings of the ACM Web Conference 2022}, pages
  2172--2182.

\bibitem[Covington et~al., 2016]{covington2016deep}
Covington, P., Adams, J., and Sargin, E. (2016).
\newblock Deep neural networks for youtube recommendations.
\newblock In {\em Proceedings of the 10th ACM conference on recommender
  systems}, pages 191--198.

\bibitem[Guo et~al., 2017]{guo2017deepfm}
Guo, H., Tang, R., Ye, Y., Li, Z., and He, X. (2017).
\newblock Deepfm: a factorization-machine based neural network for ctr
  prediction.
\newblock {\em arXiv preprint arXiv:1703.04247}.

\bibitem[He et~al., 2020]{he2020momentum}
He, K., Fan, H., Wu, Y., Xie, S., and Girshick, R. (2020).
\newblock Momentum contrast for unsupervised visual representation learning.
\newblock In {\em Proceedings of CVPR}.

\bibitem[Hidasi et~al., 2015]{gru4rec}
Hidasi, B., Karatzoglou, A., Baltrunas, L., and Tikk, D. (2015).
\newblock Session-based recommendations with recurrent neural networks.
\newblock {\em arXiv preprint arXiv:1511.06939}.

\bibitem[Jin et~al., 2020]{jin2020multi}
Jin, B., Gao, C., He, X., Jin, D., and Li, Y. (2020).
\newblock Multi-behavior recommendation with graph convolutional networks.
\newblock In {\em Proceedings of SIGIR}.

\bibitem[Kang and McAuley, 2018]{sasrec}
Kang, W.-C. and McAuley, J. (2018).
\newblock Self-attentive sequential recommendation.
\newblock In {\em 2018 IEEE international conference on data mining (ICDM)},
  pages 197--206. IEEE.

\bibitem[Li et~al., 2019]{li2019multi}
Li, C., Liu, Z., Wu, M., Xu, Y., Zhao, H., Huang, P., Kang, G., Chen, Q., Li,
  W., and Lee, D.~L. (2019).
\newblock Multi-interest network with dynamic routing for recommendation at
  tmall.
\newblock In {\em Proceedings of the 28th ACM international conference on
  information and knowledge management}, pages 2615--2623.

\bibitem[Li and Tuzhilin, 2020]{li2020ddtcdr}
Li, P. and Tuzhilin, A. (2020).
\newblock Ddtcdr: Deep dual transfer cross domain recommendation.
\newblock In {\em Proceedings of WSDM}.

\bibitem[Liu et~al., 2021]{CT4Rec}
Liu, C., Liu, X., Zheng, R., Zhang, L., Liang, X., Li, J., Wu, L., Zhang, M.,
  and Lin, L. (2021).
\newblock C$^{2}$-rec: An effective consistency constraint for sequential
  recommendation.
\newblock {\em arXiv preprint arXiv:2112.06668}.

\bibitem[Liu et~al., 2023a]{liu2023graph}
Liu, S., Meng, Z., Macdonald, C., and Ounis, I. (2023a).
\newblock Graph neural pre-training for recommendation with side information.
\newblock {\em TIS}.

\bibitem[Liu et~al., 2023b]{liu2023ufnrec}
Liu, X., Liu, C., Wang, P., Zheng, R., Zhang, L., Lin, L., Chen, Z., and Fu, L.
  (2023b).
\newblock Ufnrec: Utilizing false negative samples for sequential
  recommendation.
\newblock In {\em Proceedings of the 2023 SIAM International Conference on Data
  Mining (SDM)}, pages 46--54. SIAM.

\bibitem[Nie et~al., 2022]{nie2022mic}
Nie, P., Lu, Y., Zhang, S., Zhao, M., Xie, R., Wang, W.~Y., and Ren, Y. (2022).
\newblock Mic: Model-agnostic integrated cross-channel recommender.
\newblock In {\em Proceedings of the 31st ACM International Conference on
  Information \& Knowledge Management}, pages 3400--3409.

\bibitem[Oord et~al., 2018]{oord2018representation}
Oord, A. v.~d., Li, Y., and Vinyals, O. (2018).
\newblock Representation learning with contrastive predictive coding.
\newblock {\em arXiv preprint arXiv:1807.03748}.

\bibitem[Rendle, 2010]{rendle2010factorization}
Rendle, S. (2010).
\newblock Factorization machines.
\newblock In {\em 2010 IEEE International conference on data mining}, pages
  995--1000. IEEE.

\bibitem[Song et~al., 2019]{song2019autoint}
Song, W., Shi, C., Xiao, Z., Duan, Z., Xu, Y., Zhang, M., and Tang, J. (2019).
\newblock Autoint: Automatic feature interaction learning via self-attentive
  neural networks.
\newblock In {\em Proceedings of the 28th ACM International Conference on
  Information and Knowledge Management}, pages 1161--1170.

\bibitem[Sun et~al., 2019]{bert4rec}
Sun, F., Liu, J., Wu, J., Pei, C., Lin, X., Ou, W., and Jiang, P. (2019).
\newblock Bert4rec: Sequential recommendation with bidirectional encoder
  representations from transformer.
\newblock In {\em Proceedings of the 28th ACM international conference on
  information and knowledge management}, pages 1441--1450.

\bibitem[Wei and Zou, 2019]{wei2019eda}
Wei, J. and Zou, K. (2019).
\newblock Eda: Easy data augmentation techniques for boosting performance on
  text classification tasks.
\newblock {\em arXiv preprint arXiv:1901.11196}.

\bibitem[Wu et~al., 2022a]{wu2022disentangled}
Wu, J., Fan, W., Chen, J., Liu, S., Li, Q., and Tang, K. (2022a).
\newblock Disentangled contrastive learning for social recommendation.
\newblock In {\em Proceedings of CIKM}.

\bibitem[Wu et~al., 2022b]{wu2022multi}
Wu, Y., Xie, R., Zhu, Y., Ao, X., Chen, X., Zhang, X., Zhuang, F., Lin, L., and
  He, Q. (2022b).
\newblock Multi-view multi-behavior contrastive learning in recommendation.
\newblock In {\em Proceedings of DASFAA}.

\bibitem[Xie et~al., 2022]{xie2023multi}
Xie, R., Qiu, Z., Zhang, B., and Lin, L. (2022).
\newblock Multi-granularity item-based contrastive recommendation.
\newblock {\em arXiv preprint arXiv:2207.01387}.

\bibitem[Xu~Xie et~al., 2021]{CL4SRec}
Xu~Xie, F.~S., Liu, Z., Wu, S., Gao, J., Ding, B., and Cui, B. (2021).
\newblock Contrastive learning for sequential recommendation.

\bibitem[Yao et~al., 2021]{yao2021self}
Yao, T., Yi, X., Cheng, D.~Z., Yu, F., Chen, T., Menon, A., Hong, L., Chi,
  E.~H., Tjoa, S., Kang, J., et~al. (2021).
\newblock Self-supervised learning for large-scale item recommendations.
\newblock In {\em Proceedings of the 30th ACM International Conference on
  Information \& Knowledge Management}, pages 4321--4330.

\bibitem[Zeng et~al., 2021]{zeng2021knowledge}
Zeng, Z., Xiao, C., Yao, Y., Xie, R., Liu, Z., Lin, F., Lin, L., and Sun, M.
  (2021).
\newblock Knowledge transfer via pre-training for recommendation: A review and
  prospect.
\newblock {\em Frontiers in big Data}.

\bibitem[Zhang et~al., 2017]{zhang2017mixup}
Zhang, H., Cisse, M., Dauphin, Y.~N., and Lopez-Paz, D. (2017).
\newblock mixup: Beyond empirical risk minimization.
\newblock {\em arXiv preprint arXiv:1710.09412}.

\bibitem[Zhang et~al., 2021]{causerec}
Zhang, S., Yao, D., Zhao, Z., Chua, T.-S., and Wu, F. (2021).
\newblock Causerec: Counterfactual user sequence synthesis for sequential
  recommendation.
\newblock In {\em Proceedings of the 44th International ACM SIGIR Conference on
  Research and Development in Information Retrieval}, pages 367--377.

\bibitem[Zhou et~al., 2021]{CLRec}
Zhou, C., Ma, J., Zhang, J., Zhou, J., and Yang, H. (2021).
\newblock Contrastive learning for debiased candidate generation in large-scale
  recommender systems.
\newblock In {\em Proceedings of the 27th ACM SIGKDD Conference on Knowledge
  Discovery \& Data Mining}, pages 3985--3995.

\bibitem[Zhou et~al., 2020]{zhou2020s3}
Zhou, K., Wang, H., Zhao, W.~X., Zhu, Y., Wang, S., Zhang, F., Wang, Z., and
  Wen, J.-R. (2020).
\newblock S3-rec: Self-supervised learning for sequential recommendation with
  mutual information maximization.
\newblock In {\em CIKM}.

\bibitem[Zhu et~al., 2022]{zhu2022personalized}
Zhu, Y., Tang, Z., Liu, Y., Zhuang, F., Xie, R., Zhang, X., Lin, L., and He, Q.
  (2022).
\newblock Personalized transfer of user preferences for cross-domain
  recommendation.
\newblock In {\em Proceedings of WSDM}.

\end{thebibliography}

\end{document}